\documentclass{iaus}
\usepackage{graphicx}

\def\tex {\ifmmode{{T}_{\rm ex}}\else{$T_{\rm ex}$}\fi}
\def\tmb {\ifmmode{{T}_{\rm mb}}\else{$T_{\rm mb}$}\fi}
\def\ci     {\ifmmode{{\rm C}{\rm \small I}}\else{C\ts {\scriptsize I}}\fi}
\def\hi     {\ifmmode{{\rm H}{\rm \small I}}\else{H\ts {\scriptsize I}}\fi}
\def\hh     {\ifmmode{{\rm H}_2}\else{H$_2$}\fi}

\def\ts     {\thinspace}
\def\kms    {\ifmmode{{\rm \ts km\ts s}^{-1}}\else{\ts km\ts s$^{-1}$}\fi}
\def\msol   {\ifmmode{{\rm M}_{\odot}}\else{M$_{\odot}$}\fi}
\def\lsol   {\ifmmode{{\rm L}_{\odot}}\else{L$_{\odot}$}\fi}
\def\zsol   {\ifmmode{{\rm Z}_{\odot}}\else{Z$_{\odot}$}\fi}
\def\etal   {{\rm et\ts al.}~}

\title[Gas fraction and star formation efficiency] 
{Star Formation Efficiency at Intermediate Redshift}

\author[F. Combes et al.]   
{F. Combes$^1$,
S. Garc\'{\i}a-Burillo$^{2}$,
J. Braine$^{3}$,
E. Schinnerer$^{4}$,
F. Walter$^{4}$,
L. Colina$^{5}$
}

\affiliation{$^1$Observatoire de Paris, LERMA (CNRS:UMR8112), \\ 61 Av. de l'Observatoire, 
F-75014, Paris, France
\\ email: {\tt francoise.combes@obspm.fr} \\[\affilskip]
$^2$Observatorio Astron\'omico Nacional (OAN)-Observatorio de Madrid, \\
Alfonso XII, 3, 28014-Madrid, Spain \\[\affilskip]
$^3$Laboratoire d'Astrophysique de Bordeaux, UMR 5804,Universit\'e Bordeaux~I, \\
BP 89, 33270 Floirac, France \\[\affilskip]
$^4$Max-Planck-Institut f\"ur Astronomie (MPIA), K\"onigstuhl 17, 
69117 Heidelberg, Germany \\[\affilskip]
$^5$Departamento de Astrofisica, Centro de Astrobiologia (CSIC/INTA), \\ Torrej\'on de Ardoz, 28850 Madrid, Spain
}
\pubyear{2012}
\volume{292}  
\pagerange{119--126}
\jname{Molecular Gas, Dust, and Star Formation in Galaxies}
\editors{Tony Wong \& Juergen Ott, eds}
\begin{document}

\maketitle

\begin{abstract}
 Star formation is evolving very fast in the second half of the Universe,
and it is yet unclear whether this is due to evolving gas
content, or evolving star formation efficiency (SFE). We have carried out
a survey of ultra-luminous galaxies (ULIRG) between z=0.2 and 1, to check
the gas fraction in this domain of redshift which is still poorly known.
  Our survey with the IRAM-30m detected 33 galaxies out of 69, and
we derive a significant evolution of both the gas fraction and
SFE  of ULIRGs over the whole period, and in particular a turning point around z=0.35.
  The result is sensitive to the CO-to-\hh\, conversion factor adopted, and 
both gas fraction and SFE have comparable evolution, when we adopt the
low starburst conversion factor of $\alpha$ =0.8 M$_\odot$ (K \kms\, pc$^2$)$^{-1}$.
Adopting a higher $\alpha$ will increase the role of the gas fraction.
 Using $\alpha$ =0.8, the SFE and the gas fraction for z$\sim$0.2-1.0 ULIRGs are found to be significantly higher, 
by a factor 3, than for local ULIRGs, and are comparable to high redshift ones.  
We compare this evolution to the expected cosmic \hh\, abundance and the cosmic star formation history.
\end{abstract}

\firstsection 
\section{Introduction}
Star formation (SF) was processing at a much larger rate in galaxies in the first half of the universe history, 
and the most striking feature in the cosmic SF rate density is the decline by a factor $\sim$ 10 since z=1 
(Madau \etal 1998, Hopkins \& Beacom 2006). Several factors could be invoked to explain such a behavior: 
first the gas fraction in star forming galaxies is likely to have been higher in the past, as already suggested by 
CO surveys, tracing the molecular gas content of galaxies.  Locally, the gas fraction for giant spirals is about 7-10\% 
(Leroy  \etal 2008, Saintonge \etal 2011a), while  at z$\sim$1.2 it increases to 
34$\pm$5\% and at z$\sim$2.3 to 44$\pm$6\% (Tacconi
\etal 2010, Daddi \etal 2010). Second, the star formation efficiency might have been higher in the past, 
due to the dynamical trigger of galaxy interactions, whose frequency increases with redshift 
(e.g. Conselice \etal 2009, Kartaltepe  \etal 2010), and also the more violent instabilities in more unstable disks, 
with lower bulge-to-disk ratios. The star formation efficiency (SFE) defined as the ratio of SFR to gas content, 
has been observed to increase with redshift (e.g. Greve \etal 2005), even for the most extreme starbursts, 
represented by ultra-luminous infrared galaxies (ULIRG). This tendency is however supported mainly by 
comparing local and high-z galaxies, at z$>$1, and very little is known about the molecular gas content of galaxies 
at intermediate redshift between z=0.2 and 1. This CO desert is mainly due to observational difficulties, and 
motivated our CO survey of starburst galaxies in this redshift range. A first study at 0.2$<$z$<$0.6 
(Combes \etal 2011) has indeed confirmed a strong increase of SFE in this redshift range.

\begin{figure}[h!]
\begin{center}
\includegraphics[width=6cm]{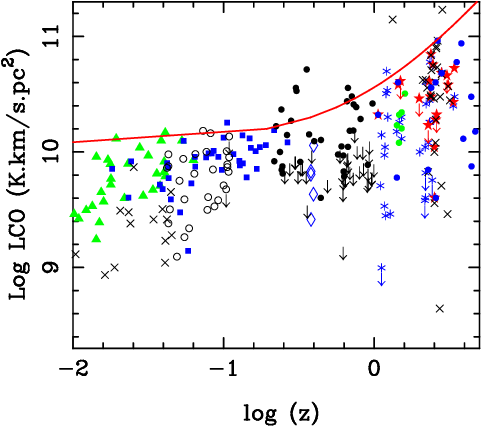} 
\includegraphics[width=6cm]{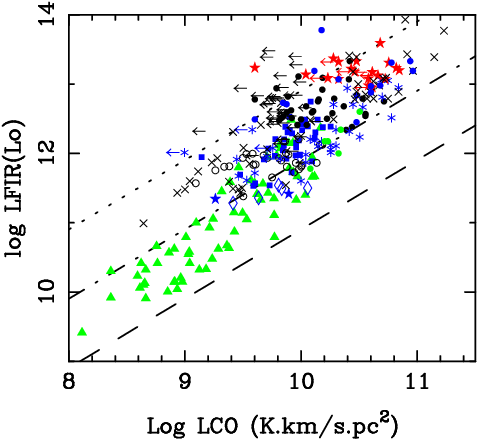} 
 \caption{{\bf Left:} CO luminosities, a proxy for total \hh\, masses, versus redshift,
for the objects of our sample
(filled black circles, and arrows as upper limits),
compared  to various data from the literature:
green triangles are from Gao \& Solomon (2004),
blue squares from Solomon \etal (1997),
open circles from Chung \etal (2009),
blue diamonds from Geach \etal (2009, 2011),
black crosses from Iono \etal (2009),
red stars, from Greve \etal (2005),
 green filled circles from Daddi \etal (2010),
blue asterisks from Genzel \etal (2010), and  blue filled circles
from Solomon \& vanden Bout (2005).
For illustration purposes only,
the red curve is the power law in (1+z)$^{1.6}$ for $\Omega_\hh$/$\Omega_\hi$
proposed by Obreschkow \& Rawlings (2009).
{\bf Right:} Correlation between far infrared and CO luminosities for
the same data.   The 3 lines are for L$_{\rm FIR}$/M(\hh)=10,
100 and 1000 \lsol/\msol\, from bottom to top, assuming a conversion factor
$\alpha$ = 0.8  M$_\odot$ (K \kms\, pc$^2$)$^{-1}$.
The three lines correspond to gas depletion time-scales of 580 (bottom), 58 (middle)
and 5.8 Myr (top).}
   \label{fig1}
\end{center}
\end{figure}

\section{Star formation efficiency and gas fraction}

We have now completed our CO survey of 69 starburst galaxies with 0.2$<$z$<$1.0, and 
obtained a global detection rate of 48\%. Since the conversion factor is a key parameter
in this study, we have obtained CO detections in several J-lines for a dozen of galaxies, and
found different excitation for the gas. When the gas mass could be derived from the dust
emission however, the resulting values conforted our adoption of the $\alpha$=0.8 
conversion factor (Combes \etal 2012). Mapping of some galaxies with the IRAM interferometer
showed that the molecular gas is extended at kpc scales, and not only confined to 
a nuclear component (Combes \etal 2006, and in prep.).

Figure \ref{fig1} shows that the data gap at intermediate redshift is now filled. The CO
luminosities of starburst galaxies display an envelope which is constantly rising
with redshift in this interval. The FIR to CO correlation is non-linear, and define
gas depletion time between 6 and 600 Myr (fig \ref{fig1} right).

\begin{figure}[h!]
\begin{center}
\includegraphics[width=6cm]{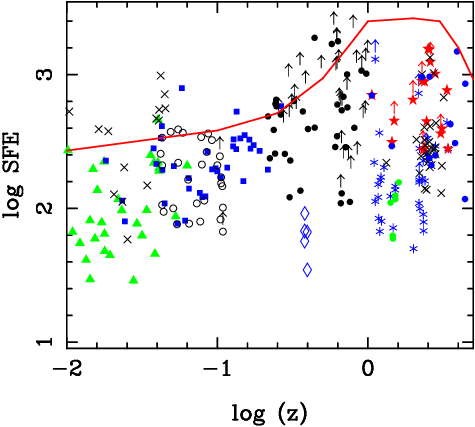} 
\includegraphics[width=6cm]{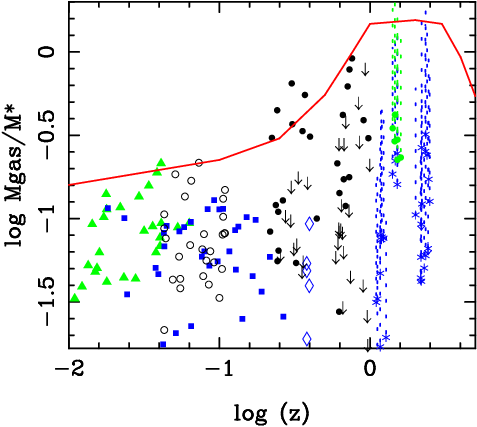} 
 \caption{{\bf Left:} The star formation efficiency versus redshift z. The symbols
are the same as in Fig \ref{fig1}.
The red curve is a schematic line summarizing the evolution of cosmic star formation density,
from the compilation by Hopkins \& Beacom (2006), complemented with 
recent work by Kistler \etal (2009) and Bouwens \etal  (2008).
This indicative curve is logarithmic and can be translated vertically.
{\bf Right:} The gas to stellar mass ratio vs redshift, with the same data.
The points of the high-z samples (blue asterisks and green dots),
have been continued by a dotted line joining the two extreme values of
gas fraction, obtained with conversion factors $\alpha$=0.8 and 4.6 M$_\odot$ (K \kms\, pc$^2$)$^{-1}$.
}
   \label{fig2}
\end{center}
\end{figure}

\begin{figure}[h!]
\begin{center}
\includegraphics[width=6cm]{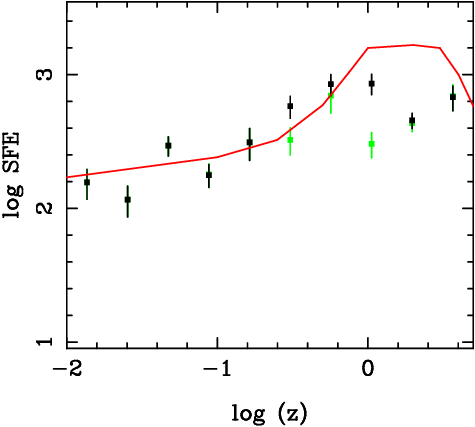} 
\includegraphics[width=6cm]{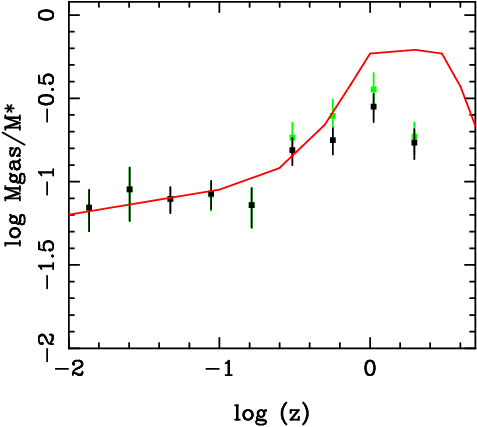} 
 \caption{Evolution with redshift of averaged quantities, SFE at {\bf left}, and gas to stellar mass ratio
at {\bf right}. The average of detected points only is plotted in green, and with the 3$\sigma$ upper limits in black
(for high-z samples only). The error bars are based on Poisson noise.
The red line is the same as in Figure  \ref{fig2}.
}
   \label{fig3}
\end{center}
\end{figure}

The gas fraction requires the determination of stellar mass, which was obtained
through SED-fitting of optical and near-infrared luminosities (Combes \etal 2012).
We define the star formation efficiency as the ratio of far infrared luminosity
(proxy of star formation rate) and the molecular mass derived from the CO luminosity,
with a constant conversion factor $\alpha$= 0.8 M$_\odot$ (K \kms\, pc$^2$)$^{-1}$.
The SFE and gas fraction are plotted versus redshift and compared to available data
in Fig  \ref{fig2}.  Both display an envelope which rises with redshift in the
intermediate range between 0.2 and 1.0, the range where the cosmic star formation
rate density is remarkably increasing by an order of magnitude.
These trends are also visible in the bin-averages of the data,
taking or not the upper limits into account (cf fig  \ref{fig3}).

\section{Conclusion}

  The increase of both star formation efficiency and gas fraction with redshift
in the intermediate range for starburst galaxies suggests that both factors
play a role in driving the evolution of the  cosmic star formation
rate density. The relative influence of each physical quantity is 
strongly linked to the CO-to-\hh\, conversion factor. If a standard Milky Way ratio
is used for these galaxies, their gas fraction will then dominate the evolution.
 We have tried to relate the star formation efficiency to the compactness 
of the starburst, measured from the half-light radius in the I-band (corresponding
to the blue band in the rest frame).  There is indeed an anticorrelation
of SFE with half-light radius, but with a large scatter.
 The SFE was also plotted with respect to the specific star formation rate
(or SFR per unit stellar mass), and compared to the locus of local ``normal'' star forming galaxies
from the COLD GASS survey (Saintonge \etal 2011b). The starburst galaxies
are globally located below, meaning that their gas content is significantly higher.


\begin{thebibliography}{}
\bibitem[]{}Bouwens R.J., Illingworth G.D., Franx M., Ford H.: 2008, \textit{ApJ} 686, 230 
\bibitem[]{}Chung, A., Narayanan, G., Yun, M. S., Heyer, M.,Erickson, N.R.: 2009, \textit{AJ} 138, 858
\bibitem[]{}Combes, F., Garc\'{\i}a-Burillo, S., Braine, J. \etal: 2006, \textit{A\&A} 460, L49 
\bibitem[]{}Combes, F., Garc\'{\i}a-Burillo, S., Braine, J. \etal: 2011, \textit{A\&A} 528, A124 
\bibitem[]{}Conselice, C. J., Yang, C., Bluck, A.F.L.: 2009, \textit{MNRAS} 394, 1956 
\bibitem[]{}Daddi E., Bournaud, F., Walter, F.\etal: 2010 \textit{ApJ} 713, 686
\bibitem[]{}Gao Y., Solomon P.M.: 2004 \textit{ApJ}S 152, 63
\bibitem[]{}Geach, J.E., Smail I., Coppin K. \etal: 2009, \textit{MNRAS} 395, L62 
\bibitem[]{}Geach, J.E., Smail I., Moran S.M. \etal: 2011, \textit{ApJ} 730, L19 
\bibitem[]{}Genzel, R., Tacconi, L. J., Gracia-Carpio, J. \etal: 2010, \textit{MNRAS} 407, 2091
\bibitem[]{}Greve. T.R., Bertoldi, F., Smail, I. \etal: 2005, \textit{MNRAS} 359, 1165
\bibitem[]{}Hopkins, A.M., Beacom J.F.: 2006, \textit{ApJ}, 651, 142 
\bibitem[]{}Iono, D., Wilson, C. D., Yun, M.S. \etal: 2009, \textit{ApJ}, 695, 1537
\bibitem[]{}Kartaltepe, J. S.,Sanders, D. B., Le Floc'h, E. \etal: 2010, \textit{ApJ} 721, 98 
\bibitem[]{}Kistler M.D., Y\"uksel H., Beacom J.F. \etal: 2009, \textit{ApJ} 705, L104 
\bibitem[]{}Leroy, A. K., Walter, F., Brinks, E. \etal 2008, \textit{AJ} 136, 2782 
\bibitem[]{}Madau P., Pozzetti L., Dickinson M.E.: 1998, \textit{ApJ} 498, 106
\bibitem[]{}Obreschkow, D., Rawlings, S.: 2009 \textit{ApJ} 696, L129 
\bibitem[]{}Saintonge, A., Kauffmann, G., Kramer, C. \etal: 2011a, \textit{MNRAS} 415, 32 
\bibitem[]{}Saintonge, A., Kauffmann, G., Wang J. \etal: 2011b, \textit{MNRAS} 415, 61 
\bibitem[]{}Solomon P., Downes D., Radford S., Barrett J.: 1997, \textit{ApJ} 478, 144
\bibitem[]{}Solomon P., Vanden Bout P.A.: 2005, \textit{ARAA}  43, 677
\bibitem[]{}Tacconi L.J., Genzel R., Neri R. \etal: 2010, \textit{Nature} 463, 781
\end{thebibliography}
\end{document}